\documentclass[journal=jacsat,manuscript=article]{achemso}

\usepackage{chemformula} 
\usepackage[T1]{fontenc} 
\usepackage{amssymb}
\usepackage{float}
\usepackage[T1]{fontenc} 
\usepackage[section]{placeins}
\usepackage{bm}

\DeclareUnicodeCharacter{2212}{\ensuremath{-}}

\title{Rotational Soft Modes and Octahedral Distortion as Design Principles for Ultralow Thermal Conductivity in Halide Materials}

\author{Yu Wu}
\affiliation{Advanced Thermal Management Technology and Functional Materials Laboratory, Ministry of Education Key Laboratory of NSLSCS, School of Energy and Mechanical Engineering, Nanjing Normal University, Nanjing 210023, P. R. China}

\author{Yufan Liu}
\affiliation{Advanced Thermal Management Technology and Functional Materials Laboratory, Ministry of Education Key Laboratory of NSLSCS, School of Energy and Mechanical Engineering, Nanjing Normal University, Nanjing 210023, P. R. China}

\author{Luman Shang}
\affiliation{Advanced Thermal Management Technology and Functional Materials Laboratory, Ministry of Education Key Laboratory of NSLSCS, School of Energy and Mechanical Engineering, Nanjing Normal University, Nanjing 210023, P. R. China}

\author{Shuming Zeng}
\email{zengsm@yzu.edu.cn}
\affiliation{College of Physics Science and Technology, Yangzhou University, Jiangsu, Yangzhou 225009, China}

\author{Liujiang Zhou}
\affiliation{School of Physics, State Key Laboratory of Electronic Thin Films and Integrated Devices, University of Electronic Science and Technology, Sichuan, Chengdu 610054, China}

\author{Hao Zhang}
\affiliation{College of Future Information Technology and Department of Optical Science and Engineering and State Key Laboratory of Photovoltaic Science and Technology, Fudan University, Shanghai 200433, China}

\author{Chenhan Liu}
\email{chenhanliu@njnu.edu.cn}
\affiliation{Advanced Thermal Management Technology and Functional Materials Laboratory, Ministry of Education Key Laboratory of NSLSCS, School of Energy and Mechanical Engineering, Nanjing Normal University, Nanjing 210023, P. R. China}

\begin{document}

\begin{abstract}
We establish that ultralow lattice thermal conductivity in halide perovskites and related octahedral framework materials arises from two distinct and complementary mechanisms: (i) halogen–halogen-enabled rotational soft modes that reshape the low-frequency spectrum and intensify phonon scattering, and (ii) static octahedral distortions that further enhance anharmonicity and reduce phonon lifetimes. Using first-principles calculations on CsPbBr$_3$, we demonstrate that Br–Br interactions induce rotational soft modes that decongest the phonon spectrum and enhance three- and four-phonon scattering, strongly suppressing particle-like thermal conductivity ($\kappa_p$). Independently, static octahedral distortions further reduce $\kappa_p$ by amplifying anharmonicity while leaving wave-like conductivity ($\kappa_c$) intact. Based on these mechanistic insights, we introduce a geometric distortion factor $\rho$ and perform a high-throughput screening that first selects materials with halogen-coordinated octahedral building blocks—ensuring the presence of rotational soft modes—and then identifies those with pronounced distortion. This strategy uncovers TaGaI$_8$ with an ultralow $\kappa_L = 0.11$ W/mK at room temperature. This work establishes halogen–halogen-enabled rotational soft modes and octahedral distortions as transferable design principles for octahedra-containing halides, spanning both extended frameworks and molecular-cluster motifs, for discovering ultralow-$\kappa_L$ materials.
\end{abstract}

\flushbottom
\maketitle

\thispagestyle{empty}

\section*{Introduction}

Halide perovskites with the general chemical formula ABX$_3$ (A = Cs$^+$, CH$_3$NH$_3^+$; B = Pb$^{2+}$, Sn$^{2+}$; X = Cl$^-$, Br$^-$, I$^-$) have emerged as a remarkable class of semiconductor materials, demonstrating exceptional optoelectronic properties that have revolutionized photovoltaic and light-emitting diode research over the past decade~\cite{He2023,Brenner2016,Green2014,Liu2020b,Isikgor2022,Haque2019,Haque2020,Liang2017}. Beyond their well-established applications in solar cells and LEDs, these materials have recently attracted considerable attention for their unique thermal transport characteristics, particularly their intrinsically ultralow lattice thermal conductivity ($\kappa_{\rm L}$)~\cite{Yuan2017,Ha2015,Pazos-Outon2016,Lee2017,Zheng2024,Acharyya2022,Wu2024,Acharyya2023}. This combination of efficient charge transport and suppressed heat conduction positions halide perovskites as promising candidates for thermoelectric energy conversion and thermal barrier coating applications~\cite{Hu2021,Liu2019a,Xie2020,Haque2020b,Wang2022}.

A key structural feature of halide perovskites is the three-dimensional framework of corner-sharing BX$_6$ octahedra. As systematically elucidated by Bechtel and Van der Ven~\cite{Bechtel2018}, inorganic CsMX$_3$ perovskites (M = Pb, Sn; X = Br, I) exhibit intrinsic structural instabilities originating from cooperative tilting motions of the octahedra. These tilt modes can be classified as in-phase or out-of-phase rotations, which couple strongly with macroscopic strains and A-site displacements, ultimately stabilizing the orthorhombic $Pnma$ ground state observed in materials such as CsPbI$_3$ and CsSnI$_3$~\cite{Bechtel2018}. Importantly, the energy landscape associated with these tilt degrees of freedom exhibits pronounced anharmonicity, with the cubic and tetragonal phases residing at local maxima or saddle points rather than true minima---a direct manifestation of dynamic instabilities that persist to high temperatures~\cite{Bechtel2018,Wang2025}.

The connection between octahedral tilting and thermal transport has been corroborated by multiple studies. Acharyya et al.~\cite{Acharyya2020} demonstrated that the two-dimensional Ruddlesden-Popper perovskite Cs$_2$PbI$_2$Cl$_2$ exhibits ultralow $\kappa_{\rm L}$ ($\sim$0.37 W/mK) originating from soft optical phonon modes ($\sim$12--55 cm$^{-1}$) associated with dynamic octahedral rotations and their strong coupling with acoustic phonons. Xie and Zhao~\cite{Xie2024} further identified cation off-centering behavior---driven by $ns^2$ lone pair electrons, weak $sd^3$ orbital hybridization, or oversized coordination environments---as a root cause of ultralow thermal conductivity in diverse material families, with the core mechanism being low-frequency optical modes scattering heat-carrying acoustic phonons through acoustic--optical coupling.

Beyond the octahedral framework itself, the role of A-site cations has also sparked extensive discussion. The conventional ``rattling'' picture, which treats weakly bound guest ions as independent vibration centers that resonantly scatter phonons to reduce thermal conductivity, has been widely invoked for cage-like compounds such as skutterudites and clathrates~\cite{Christensen2008,Tadano2015} and extended to halide perovskites. However, Thakur and Giri~\cite{Thakur2023} challenged this interpretation through systematic molecular dynamics simulations on CsPbI$_3$. By comparing the thermal transport properties of the fully filled structure with an empty PbI$_6$ framework (with Cs$^+$ ions removed), they demonstrated that the addition of Cs$^+$ actually enhances thermal conductivity through vibrational hardening of the framework, with Cs$^+$ vibrations exhibiting well-defined phase relations with the framework rather than acting as independent rattlers~\cite{Thakur2023,Wu2025}. This finding further reinforces that the octahedral framework, rather than the A-site cation, is the primary determinant of thermal transport in these materials.

Recent advances have underscored the importance of accurately capturing anharmonic effects in these materials. Pandey \textit{et al.}~\cite{PANDEY2022} demonstrated that temperature-dependent phonon renormalization, particularly the hardening of low-energy optical modes associated with SnI$_6$ octahedral rotations, is essential for quantitatively predicting the lattice thermal conductivity in tin-based halide perovskites. Concurrently, our recent work on Cs$_3$Bi$_2$I$_6$Cl$_3$ revealed the electronic origin of dynamic rotations, discovering that they are driven by electrostatic repulsion of $p$-band electrons rather than the traditionally assumed $s$-band lone pairs, with ultralow $\kappa_{\rm L}$ ($<$0.2 W/mK) containing a significant glass-like thermal transport contribution~\cite{Wu2024}. Zeng \textit{et al.}~\cite{Zeng2025} further elucidated this glass-like behavior, showing that severe lattice distortion at low temperatures, driven by atomic size mismatch and shallow double-well potentials, leads to structural disorder that dominates heat conduction over anharmonic scattering. However, the specific role of static octahedral distortions—as distinct from dynamic rotations or low-temperature disorder—in modulating the two fundamental channels of heat transport, namely particle-like propagation ($\kappa_p$) and wave-like tunneling ($\kappa_c$), remains to be systematically established.

Building on this background, the present work systematically investigates the effects of octahedral rotational soft modes and static distortions on thermal transport in CsPbBr$_3$ through first-principles calculations. By selectively weakening the Br--Br interaction, we reveal the microscopic mechanism by which halogen--halogen coupling drives rotational soft modes; by introducing Pb displacements to modulate octahedral distortion, we elucidate the influence of distortion on thermal conductivity components ($\kappa_{\rm p}$ and $\kappa_{\rm c}$), phonon lifetimes, scattering phase space, and Grüneisen parameters. Based on these mechanistic insights, we develop a geometric distortion factor $\rho$ as a descriptor and apply it to high-throughput screening of crystal structure databases, identifying a family of halogen-rich compounds built from distorted halogen-coordinated octahedra, including NbAlI$_8$, NbGaI$_8$, TaAlI$_8$, and TaGaI$_8$. Despite their common 0D cluster-like building blocks, they still support low-frequency octahedral rotational modes with strong anharmonicity, underscoring the generality of the proposed principles beyond strictly corner-sharing perovskite frameworks. Detailed thermal transport analysis of TaGaI$_8$ reveals ultralow $\kappa_{\rm L}$ = 0.11 W/mK at room temperature with significant four-phonon scattering contributions. This work demonstrates that both octahedral rotational soft modes and static distortions can serve as effective strategies for tuning thermal conductivity, providing a theoretical framework for the rational design of novel low-thermal-conductivity materials.


\section*{Results and Discussion}

Figure 1 elucidates the contribution of octahedral rotational soft modes to the ultralow lattice thermal conductivity in CsPbBr$_3$. To quantitatively identify this contribution, we employed a controlled perturbation by selectively weakening the adjacent Br–Br interaction by half, and systematically compared its impact on the phonon spectra and thermal transport properties.

\textbf{Figure 1a} presents a comparison of the phonon spectra before and after weakening the Br–Br interaction at 600 K. In the pristine system, the low-frequency region exhibits clear characteristics of soft modes associated with octahedral rotations, manifesting as extremely low-frequency modes at the Brillouin zone boundaries. Upon weakening the Br–Br interaction by half, these low-frequency soft modes shift upward and undergo significant hardening. This seemingly counterintuitive trend—mode hardening upon weakening a specific interaction—highlights that the Br–Br Coulomb/short-range coupling provides a key softening (destabilizing) contribution to the rotational degrees of freedom. The physical picture can be understood as follows: the low-frequency rotational modes primarily correspond to collective lateral displacements of the halogen sublattice (the Br network) and octahedral tilting. The interactions between neighboring Br atoms are highly sensitive to such collective rotations: when a local Br undergoes a lateral displacement, the Br–Br coupling efficiently transfers this perturbation to adjacent Br atoms and activates a cooperative framework response over an extended spatial range. This Br–Br-mediated cooperativity effectively flattens the potential-energy landscape along the tilting-mode coordinate, i.e., it reduces the net harmonic stiffness (the curvature of the total energy projected onto the tilting coordinate), thereby rendering the rotational degrees of freedom as low-frequency soft modes~\cite{Bechtel2018}. When we intentionally weaken the Br–Br interaction, this destabilizing/softening contribution is reduced; as a result, the net curvature of the total potential energy along the collective tilting coordinate becomes larger (even though the Br–Br interaction itself is weakened), leading to an upshift (hardening) of the rotational mode frequencies.

Concurrent with the hardening of rotational soft modes, the lattice thermal conductivity of the system undergoes substantial changes. Notably, $\kappa_p$ increases from 0.28\,W/mK in the pristine system to 0.56\,W/mK after weakening, nearly doubling, while $\kappa_c$ remains almost unchanged. This indicates that rotational soft modes suppress thermal transport primarily through the quasiparticle channel. To further pinpoint the origin of the $\kappa_p$ increase, we employed a physical quantity substitution analysis: within the ``pre-weakening'' (pristine) framework, we individually substituted the heat capacity ($c$), group velocity ($v$), and phonon lifetime ($\tau$) with their post-weakening values to recalculate $\kappa_p$ in \textbf{Figure 1b}. The results show that replace-$c$ yields almost identical $\kappa_p$, replace-$v$ increases $\kappa_p$ by only about 17\,\%, while replace-$\tau$ boosts $\kappa_p$ by approximately 75\,\%. This clearly demonstrates that the primary source of the increased thermal conductivity after weakening the Br–Br interaction is the significant prolongation of phonon lifetimes, rather than changes in heat capacity or dispersion effects.

\textbf{Figures 1c--1f} further reveal the microscopic origins of these changes. The comparison of group velocities (\textbf{Figure 1c}) shows that after weakening the Br–Br interaction, the group velocities of modes around 0.3\,THz increase significantly due to the notable hardening of the phonon spectrum in that region. Furthermore, the hardened phonon spectrum exhibits a more concentrated distribution within the 0.5--1.5\,THz range. Governed by energy and momentum conservation, this spectral ``congestion'' profoundly affects the phonon scattering phase space. \textbf{Figures 1d} and \textbf{1e} present the changes in the weighted three-phonon phase space (WP$_3$) and weighted four-phonon phase space (WP$_4$), respectively---both show a substantial reduction within the 0.5--1.5\,THz frequency range. This suggests that in the pristine system, the Br–Br interaction drives a ``decongestion'' ((i.e., spectral spreading / reduced spectral crowding)) of the phonon spectrum: the soft modes disperse the spectrum across a wider frequency range, thereby broadening the scattering channels that satisfy energy-momentum conservation. Conversely, weakening the Br–Br interaction leads to a re-congestion of the spectrum, restricting scattering channels and reducing the phase space volume. This picture aligns perfectly with the evolution of the Grüneisen parameters. \textbf{Figure 1f} shows that weakening the Br–Br interaction significantly reduces the Grüneisen parameters, with the maximum value decreasing by nearly a factor of seven, providing direct evidence for the attenuation of anharmonic interactions.

Collectively, the systematic analysis in \textbf{Figure 1} establishes a clear causal chain: Br–Br interactions, by driving octahedral rotational soft modes, induce a ``decongestion'' effect on the phonon spectrum. This decongestion broadens the scattering channels satisfying energy-momentum conservation, enhances both the weighted three- and four-phonon phase spaces, and is accompanied by large Gr\"{u}neisen parameters (strong anharmonicity). These factors collectively result in significantly shortened phonon lifetimes and a strongly suppressed particle-like thermal conductivity $\kappa_p$. This mechanism underscores the central role of adjacent halogen interactions in regulating octahedral tilting dynamics and their consequent contribution to achieving ultralow thermal conductivity.

\begin{figure*}[ht!]
\centering
\includegraphics[width=1\linewidth]{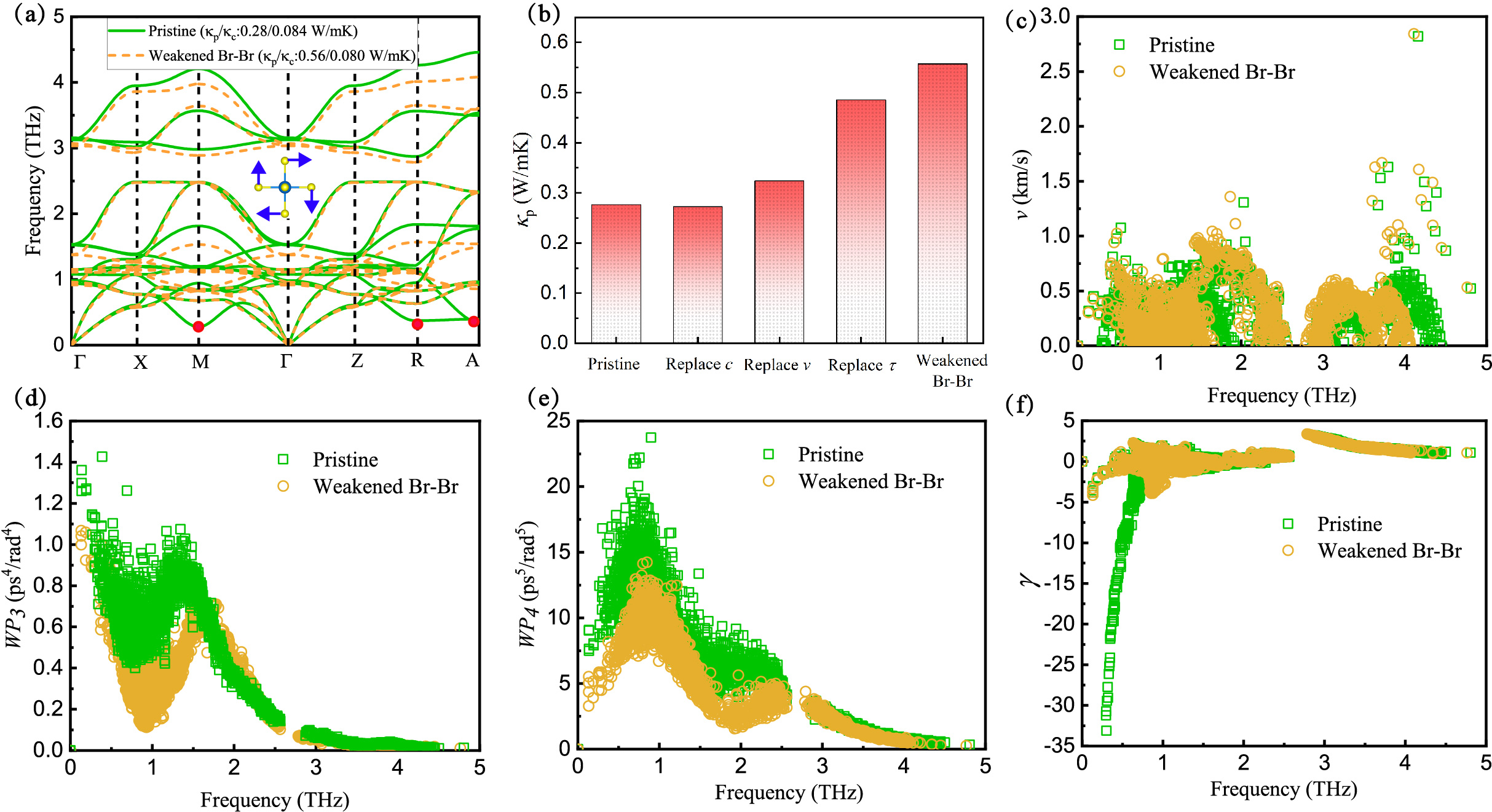}
\caption{  (a) Phonon spectra of CsPbBr$_3$ comparing the pristine system with the case after weakening Br–Br interactions by half at 600 K. The low-frequency rotational soft modes exhibit significant hardening upon weakening the interaction. (b) Substitution analysis of particle-like thermal conductivity ($\kappa_p$) performed on the pristine system by individually replacing its heat capacity ($c$), group velocity ($v$), and phonon lifetime ($\tau$) with the corresponding values obtained from the weakened system. (c) Comparison of group velocities before and after weakening. (d) Weighted three-phonon phase space (WP$_3$) before and after weakening. (e) Weighted four-phonon phase space (WP$_4$) before and after weakening. (f) Gr\"{u}neisen parameters before and after weakening.}
\label{Fig1}
\end{figure*}

\textbf{Figure 2} investigates the impact of octahedral distortion on the lattice thermal conductivity of CsPbBr$_3$, introduced by displacing the central Pb atom along the $x$ direction relative to the lattice constant. \textbf{Figure 2a} presents the thermal conductivity components ($\kappa_p$, $\kappa_c$, and total $\kappa_L = \kappa_p + \kappa_c$) as functions of the Pb displacement magnitude. While $\kappa_p$ shows a slight increase at a displacement of 0.02, it exhibits an overall decreasing trend with increasing distortion, whereas $\kappa_c$ remains largely insensitive to the displacement magnitude. At a displacement of 0.05, the total lattice thermal conductivity $\kappa_L$ decreases by approximately 20\% compared to the pristine system, indicating that large octahedral distortions can effectively reduce the lattice thermal conductivity by suppressing $\kappa_p$ without significantly enhancing the wave-like contribution $\kappa_c$. To understand the origin of the $\kappa_p$ variation, particularly the slight increase at 0.02 displacement, \textbf{Figure 2b} presents a substitution analysis performed on the pristine system by individually replacing its heat capacity ($C_v$), group velocity ($v$), and phonon lifetime ($\tau$) with the corresponding values obtained from the system with 0.05 displacement. The results show that replace-$C_v$ yields almost identical $\kappa_p$, replace-$v$ increases $\kappa_p$ by approximately 6\,\%, while replace-$\tau$ decreases $\kappa_p$ by about 29\,\%. The net suppression of $\kappa_p$ at large distortions is therefore dominated by the reduction in phonon lifetime, while the slight increase at 0.02 displacement likely arises from a competitive interplay between the modest $\tau$ reduction and the $v$ increase at small distortions.

Unlike the Br--Br-driven rotational soft modes that primarily reshape the ultralow-frequency sector, the static Pb off-centering enhances branch mixing and dispersion broadening over a wider frequency window, thereby increasing scattering channels. \textbf{Figure 2c} compares the phonon spectra of the pristine system and the system with 0.05 Pb displacement. The distorted system exhibits a pronounced ``decongestion'' of the phonon spectrum, particularly in the low-frequency region, where previously degenerate or closely clustered branches become separated and redistributed over a wider frequency range. Meanwhile, the phonon dispersion above 2\,THz becomes significantly broader than the pristine case, indicating enhanced acoustic--optical coupling. Consistent with these spectral changes, the weighted three-phonon phase space (WP$_3$) shown in \textbf{Figure 2d} increases substantially in the 0--1\,THz frequency range after introducing the 0.05 distortion, reflecting an enlarged phase space available for three-phonon scattering. This directly contributes to enhanced phonon scattering rates and consequently reduced phonon lifetimes ($\tau$). \textbf{Figure 2e} displays the weighted four-phonon phase space (WP$_4$), which also shows an increase across the entire frequency range relative to the pristine system, indicating that the distorted structure also facilitates four-phonon scattering, albeit to a lesser extent than three-phonon processes. \textbf{Figure 2f} presents the Grüneisen parameters before and after distortion, showing that octahedral distortion significantly increases the Grüneisen parameters in the low-frequency region, providing direct evidence for enhanced intrinsic anharmonicity that amplifies phonon scattering beyond what the harmonic-phase-space considerations alone would predict.

Together, these results establish that octahedral distortion serves as an effective strategy for reducing lattice thermal conductivity in halide perovskites by simultaneously modifying both harmonic properties (phonon spectrum decongestion and increased scattering phase space) and anharmonicity (elevated Gr\"{u}neisen parameters). These microscopic changes collectively result in reduced phonon lifetimes and a 20\% reduction in $\kappa_L$ at 0.05 displacement, while leaving $\kappa_c$ largely unaffected. The competitive interplay between group velocity increase and lifetime reduction at small distortions explains the non-monotonic behavior of $\kappa_p$, with the lifetime effect ultimately dominating at large distortions to achieve ultralow thermal conductivity

\begin{figure*}[ht!]
\centering
\includegraphics[width=1\linewidth]{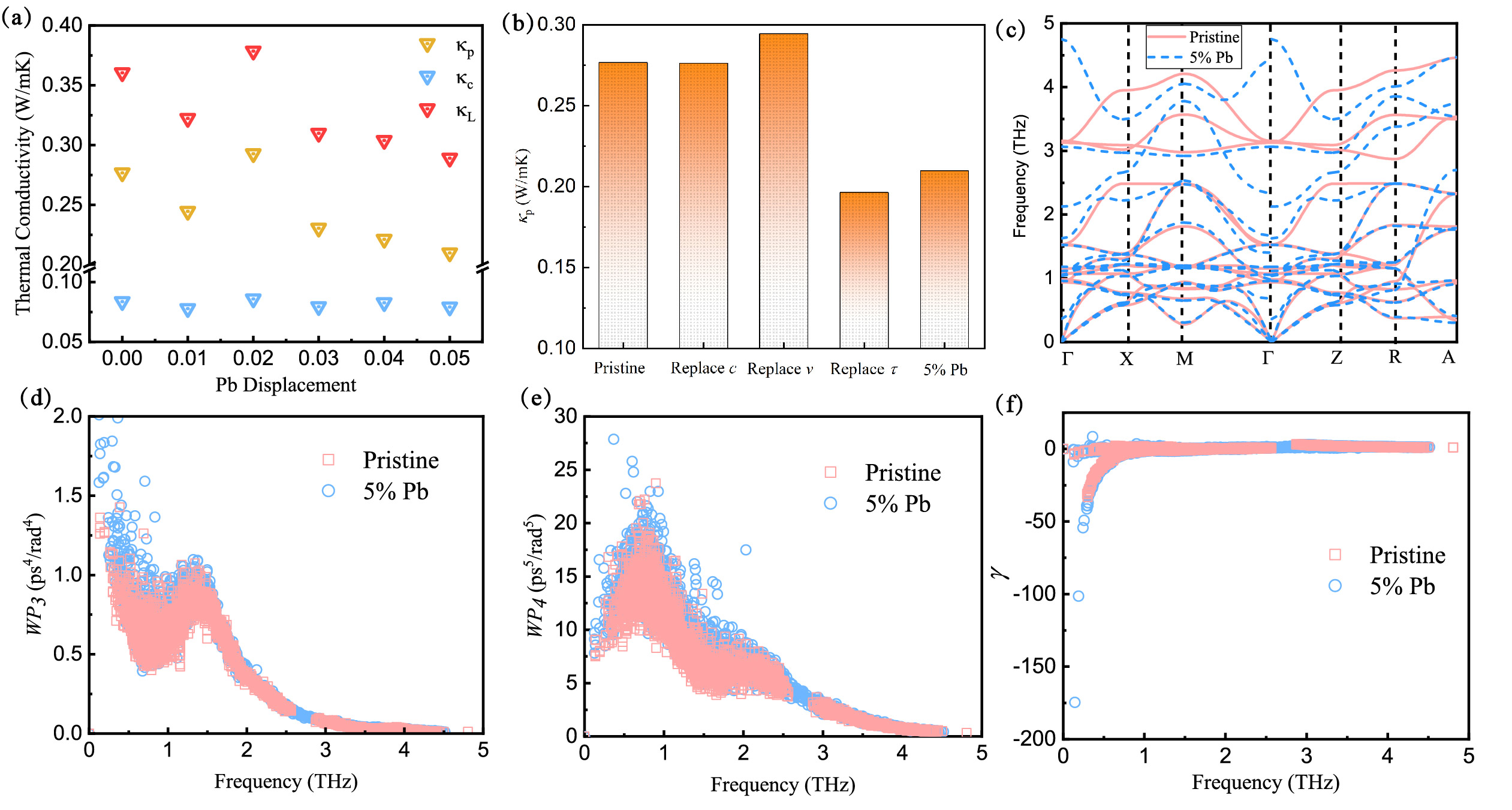}
\caption{(a) Thermal conductivity components ($\kappa_p$, $\kappa_c$, and total $\kappa_L = \kappa_p + \kappa_c$) as functions of Pb displacement magnitude along the $x$ direction relative to the lattice constant. (b) Substitution analysis of $\kappa_p$ performed on the pristine system by individually replacing its heat capacity ($c$), group velocity ($v$), and phonon lifetime ($\tau$) with the corresponding values obtained from the system with 0.05 displacement. (c) Comparison of phonon spectra between the pristine system and the system with 0.05 Pb displacement. (d) Weighted three-phonon phase space (WP$_3$) before and after distortion. (e) Weighted four-phonon phase space (WP$_4$) before and after distortion. (f) Gr\"{u}neisen parameters before and after distortion.}
\label{Fig2}
\end{figure*}

\textbf{Figure 3} presents a schematic summary illustrating how halogen-atom interaction-driven rotational soft modes and octahedral distortions collaboratively govern the ultralow lattice thermal conductivity in halide perovskites.

\begin{figure*}[ht!]
\centering
\includegraphics[width=1\linewidth]{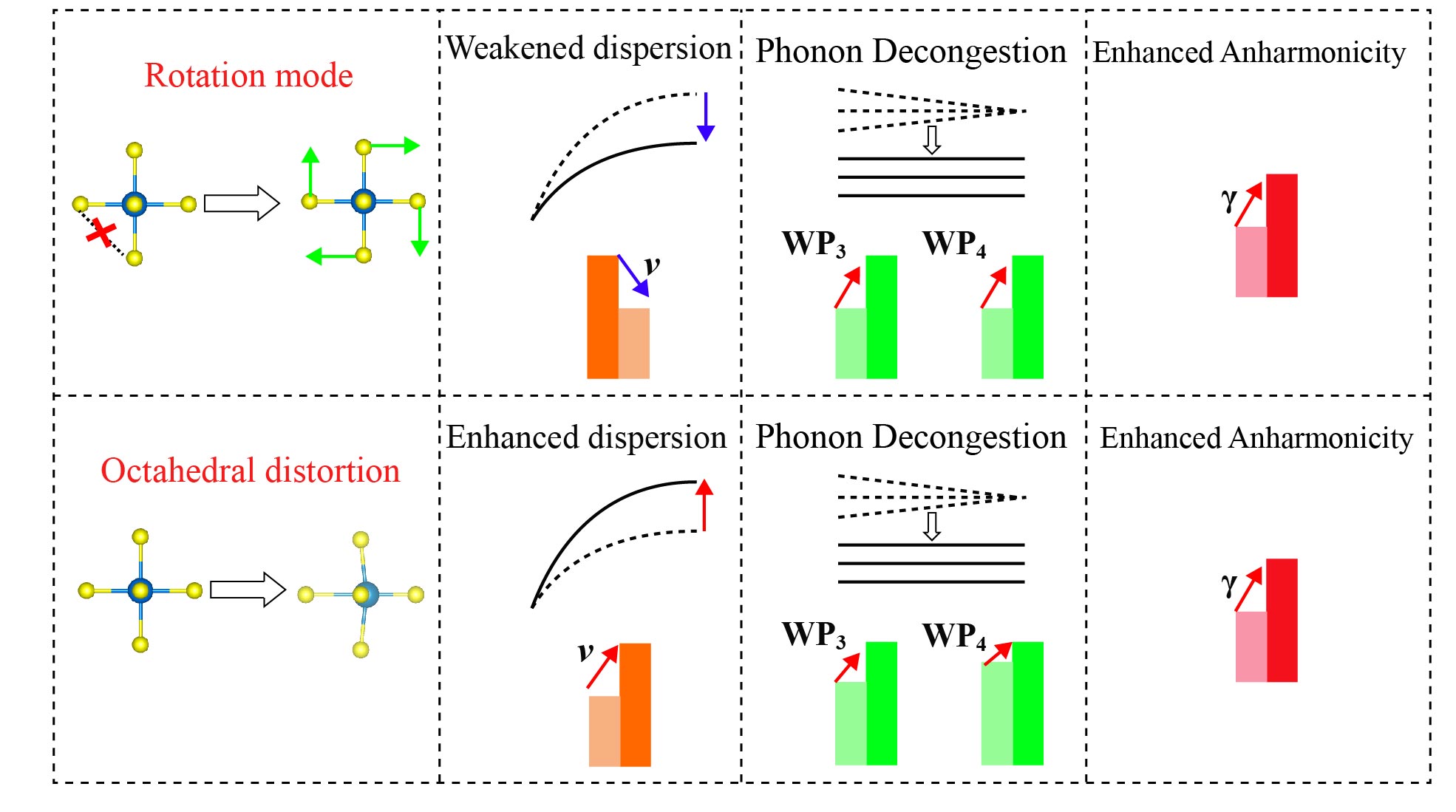}
\caption{Schematic illustration of the dual mechanisms governing ultralow lattice thermal conductivity in halide perovskites: halogen-atom interaction-driven rotational soft modes and octahedral distortions.}
\label{Fig3}
\end{figure*}

\begin{table*}[htbp]
\centering
\caption{Selected ultralow thermal conductivity materials with their chemical formulas, space group symmetries, $\kappa_p$ (W/mK), $\kappa_c$ (W/mK), $\kappa_L$ (W/mK), and distortion factor $\rho$.}
\label{tab:ultralow_kappa}
\begin{tabular}{lcccccc}
\hline
\textbf{Material} & \textbf{Symmetry} & $\bm{\kappa_p}$ & $\bm{\kappa_c}$ & $\bm{\kappa_L}$ & $\bm{\rho}$ \\
\hline
AlSbI$_6$  & $C2/m$       & 0.16 & 0.04 & 0.20 & 0.69 \\
NbAlI$_8$  & $Cmcm$       & 0.09 & 0.05 & 0.14 & 0.56 \\
NbGaI$_8$  & $Cmcm$       & 0.12 & 0.03 & 0.15 & 0.55 \\
TaAlI$_8$  & $Cmcm$       & 0.08 & 0.03 & 0.11 & 0.52 \\
TaGaI$_8$  & $Cmcm$       & 0.08 & 0.03 & 0.11 & 0.50 \\
HfI$_4$  & $C2/c$       & 0.18 & 0.04 & 0.22 & 0.41 \\
\hline
\end{tabular}
\end{table*}

The mechanistic insights gained from CsPbBr$_3$ demonstrate that both halogen-atom interaction-driven rotational soft modes and octahedral distortions are effective strategies for suppressing lattice thermal conductivity. These findings motivate a broader search for materials exhibiting similar structural features. Specifically, we target compounds containing octahedral building units with halogen ligands, where the combination of soft-mode dynamics and static octahedral distortions can cooperatively yield ultralow $\kappa_L$.

To identify candidate materials, we developed an automated high-throughput screening workflow based on geometric analysis of crystal structures. The workflow processes structure files obtained from the Materials Project database and evaluates each candidate against a set of criteria designed to capture the essential features identified in our mechanistic study. First, for each structure, we identify potential octahedral centers from a predefined set of common octahedral-forming elements (e.g., Pb, Sn, Bi, Sb, transition metals). For each candidate center, the six nearest neighbors within 4.0 \AA$\;$are taken as ligands, and all six ligands are required to be halogens (Cl/Br/I) to ensure strong halogen–halogen interactions analogous to those in CsPbBr$_3$. The six ligands must then form a geometry approximating an octahedron, assessed by checking that the angles between adjacent ligands and the central atom fall within $90^\circ \pm 25^\circ$ and that at least one angle falls within $180^\circ \pm 25^\circ$, ensuring the presence of opposing ligands necessary for a closed octahedral cage.

For each validated octahedron, we quantify its deviation from ideal octahedral geometry using a distortion factor $\rho$ that captures both bond length and angle variations:
\begin{equation}
\rho = \alpha \cdot \min\left(\frac{\sigma_d/\bar{d}}{0.1}, 1\right) + \beta \cdot \min\left(\frac{\sqrt{\frac{1}{N}\sum_{i=1}^{N} \min(|\theta_i - 90^\circ|, |\theta_i - 180^\circ|)^2}}{20^\circ}, 1\right)
\end{equation}
where $\bar{d}$ and $\sigma_d$ are the mean and standard deviation of the six bond lengths, $\theta_i$ are the 15 inter-ligand angles, and $\alpha = 0.4$, $\beta = 0.6$ are weighting coefficients. The factors $0.1$ and $20^\circ$ in the denominators normalize the bond-length and angle contributions to a 0--1 scale based on typical maximum distortions observed in halide octahedra. The normalization constants ($0.1$ and $20^\circ$ ) are chosen according to the upper-percentile statistics of bond-length and angle distortions in halogen-coordinated octahedra within the screened dataset, ensuring a robust 0–1 scaling. The factor $\rho$ ranges from 0 (perfect octahedron) to 1 (severely distorted). Finally, to ensure the octahedral framework exhibits connectivity conducive to soft-mode dynamics, we enforce that each ligand atom is shared by at most two octahedra. This constraint excludes structures where halogen atoms are over-coordinated in a manner that precludes the cooperative tilting motions essential for the low-$\kappa_L$ mechanisms identified in our study.

\begin{figure*}[ht!]
\centering
\includegraphics[width=1\linewidth]{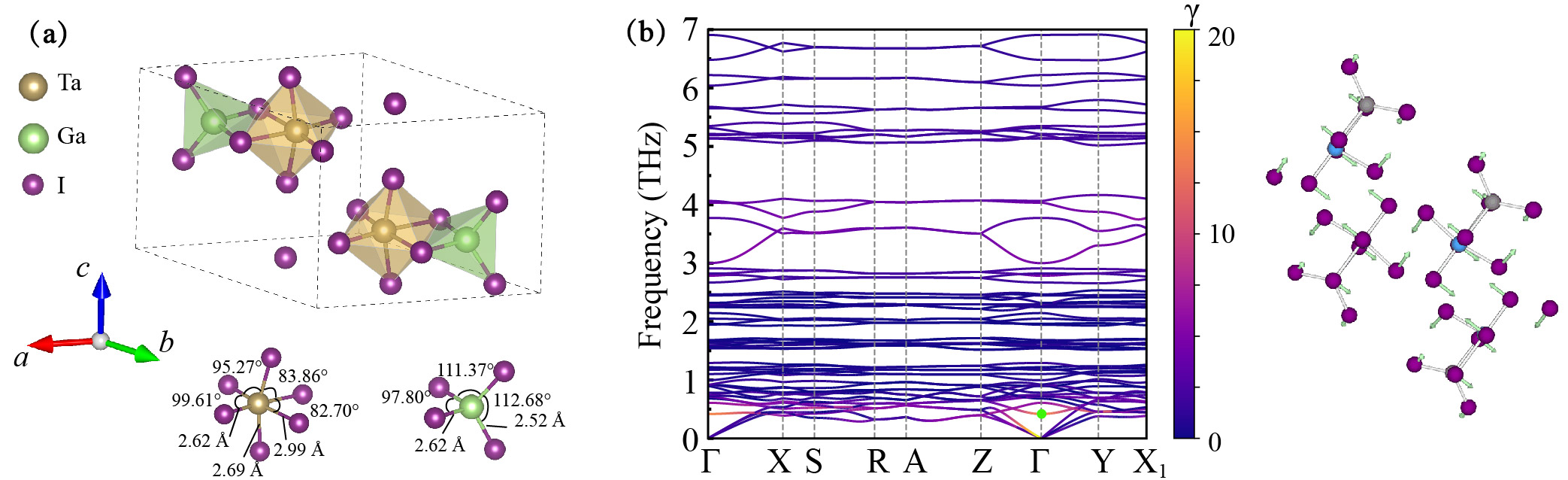}
\caption{(a) Crystal structure of TaGaI$_8$, featuring a zero-dimensional (0D) molecular motif composed of isolated clusters, each containing a distorted TaI$_6$ octahedron and a GaI$_4$ tetrahedron sharing two iodine atoms. (b) Phonon spectrum of TaGaI$_8$ projected by the Gr\"{u}neisen parameter, revealing low-frequency soft modes with strong anharmonicity.}
\label{Fig1}
\end{figure*}

The high-throughput screening workflow successfully identified several candidate materials with pronounced octahedral distortion, including four compounds—NbAlI$_8$, NbGaI$_8$, TaAlI$_8$, and TaGaI$_8$—all crystallizing in the orthorhombic space group $Cmcm$ and exhibiting large distortion factors. Among these, TaGaI$_8$ displays the lowest lattice thermal conductivity ($\kappa_L = 0.11$,W/mK) and is therefore selected for detailed structural and thermal transport analysis.

\textbf{Figure 4a} illustrates the crystal structure of TaGaI$_8$, which features a zero-dimensional (0D) molecular motif composed of isolated clusters. Each cluster consists of a TaI$_6$ octahedron and a GaI$_4$ tetrahedron that share two iodine atoms, forming a compact building unit. Within the TaI$_6$ octahedron, the Ta–I bond lengths range from 2.69 to 2.99\,\AA, with a maximum difference of 0.3\,\AA, indicating a notable distortion from ideal octahedral geometry. The I–Ta–I bond angles vary between 82.70$^\circ$ and 99.61$^\circ$, with a maximum deviation of 16.91$^\circ$ from the ideal 90$^\circ$ angle, further confirming the substantial distortion of the octahedral unit. This distortion is quantitatively captured by the distortion factor $\rho = 0.50$, consistent with the substantial deviation from perfect octahedral symmetry. Notably, TaGaI$_8$ satisfies both screening criteria: it contains halogen-coordinated octahedral building units that can support rotational soft modes, and its large $\rho$ value reflects the pronounced static distortion that further enhances anharmonicity.

\textbf{Figure 4b} presents the phonon spectrum of TaGaI$_8$ projected by the Grüneisen parameter, which measures the degree of anharmonicity for each phonon mode. Similar to CsPbBr$_3$, the low-frequency region exhibits pronounced soft-mode characteristics associated with octahedral rotations, accompanied by large Gr\"{u}neisen parameters. This indicates strong anharmonicity in these rotational modes, a key feature enabling ultralow thermal conductivity. The coexistence of significant octahedral distortion and soft-mode-driven anharmonicity in TaGaI$_8$ mirrors the dual mechanisms identified in CsPbBr$_3$, validating the screening criteria and underscoring the transferability of the design principles established in this work.

\begin{figure*}[ht!]
\centering
\includegraphics[width=1\linewidth]{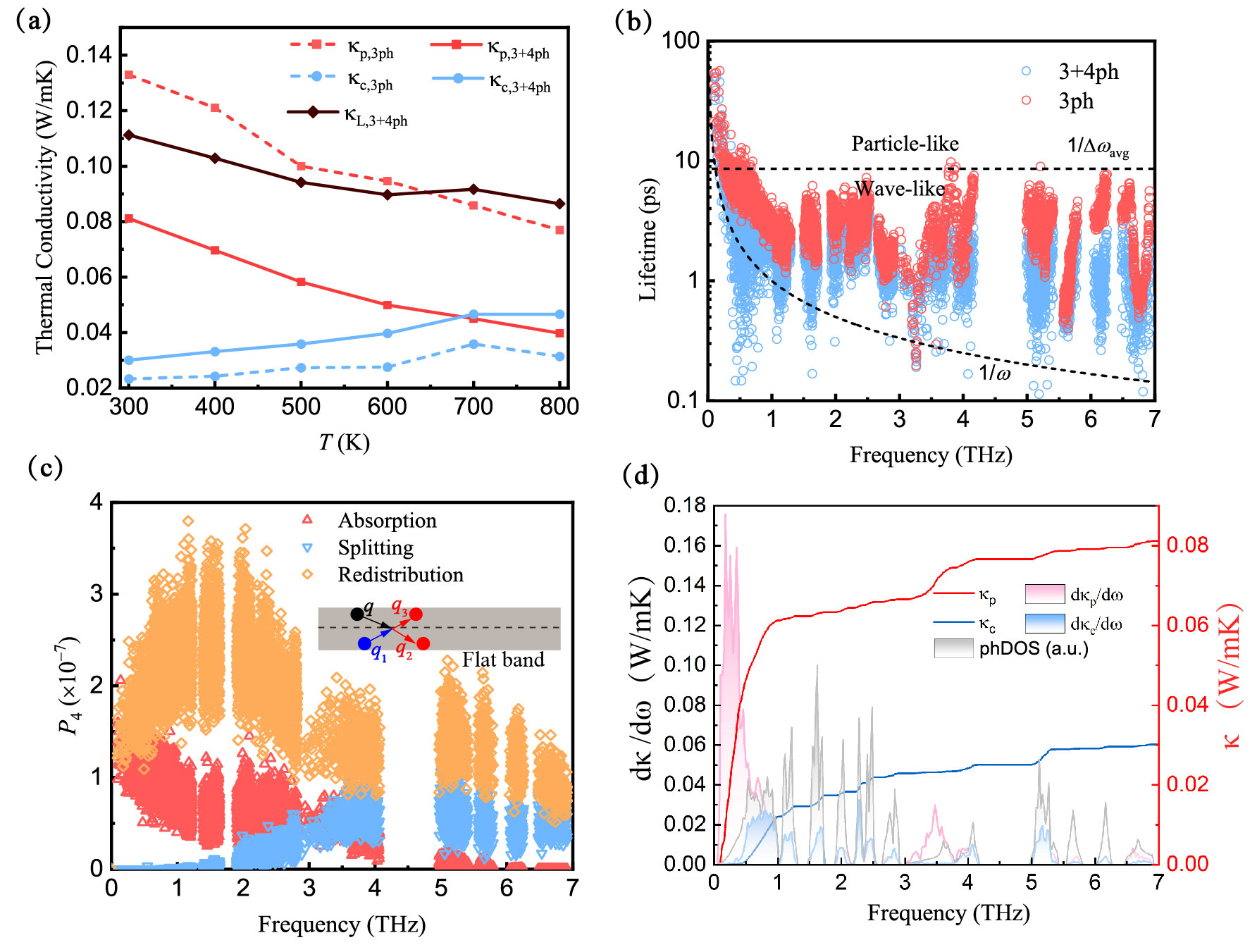}
\caption{(a) Temperature-dependent thermal conductivity components ($\kappa_p$, $\kappa_c$, and total $\kappa_L$) of TaGaI$_8$ calculated with three-phonon (3ph) and three-plus-four-phonon (3+4ph) scattering. (b) Phonon lifetimes at 300 K comparing 3ph and 3+4ph scattering. Dashed lines indicate the Ioffe-Regel ($\tau = \omega^{-1}$) and Wigner ($\tau = \Delta \omega_{\text{avg}}^{-1}$) limits separating particle-like and wave-like transport regimes.$^{22}$ (c) Four-phonon scattering phase space ($P_4$) decomposed into absorption, splitting, and redistribution processes. (d) Cumulative and differential contributions to $\kappa_p$ and $\kappa_c$ at 300 K (3+4ph).}
\label{Fig5}
\end{figure*}

We now turn to the thermal transport properties of TaGaI$_8$, the candidate with the lowest $\kappa_L$ among the screened materials. \textbf{Figure 5a} presents the temperature-dependent thermal conductivity components ($\kappa_p$, $\kappa_c$, and total $\kappa_L = \kappa_p + \kappa_c$) calculated with both three-phonon (3ph) and three-plus-four-phonon (3+4ph) scattering. At 300 K, inclusion of four-phonon scattering reduces $\kappa_p$ from 0.13 to 0.08 W/mK, a substantial decrease of 39\%, demonstrating the critical role of higher-order anharmonic processes in this material. The wave-like thermal conductivity $\kappa_c$ at 300 K increases from 0.02 to 0.03 W/mK upon including four-phonon scattering, raising its contribution to total $\kappa_L$ from 15\% to 27\%. As temperature increases, $\kappa_p$ decreases monotonically due to enhanced phonon-phonon scattering, while $\kappa_c$ exhibits a gradual increase, reaching 0.05 W/mK at 800 K under 3+4ph scattering, where it constitutes 54\% of the total $\kappa_L$. Notably, the temperature dependence of $\kappa_L$ becomes significantly weaker after including the wave-like contribution—the reduction from 300 K to 800 K is only 22\% for $\kappa_L$ (3+4ph) compared to 51\% for $\kappa_p$ (3+4ph)—indicating that wave-like tunneling processes partially offset the decline in particle-like transport at elevated temperatures.

\textbf{Figure 5b} compares the phonon lifetimes calculated with 3ph and 3+4ph scattering at 300 K. The inclusion of four-phonon scattering substantially reduces lifetimes across the entire frequency range, with the most pronounced suppression occurring in the low-to-mid frequency region. The dashed lines indicate two characteristic limits that separate different heat transport regimes: the Ioffe-Regel limit ($\tau = \omega^{-1}$) and the Wigner limit ($\tau = \Delta \omega_{\text{avg}}^{-1}$). Phonons with lifetimes above the Wigner limit propagate as well-defined particles ($\kappa_p$ contribution), while those with lifetimes between the two limits exhibit wave-like tunneling behavior ($\kappa_c$ contribution)\cite{DiLucente2023,Simoncelli2019,Wu2023b,Wu2025b}. Upon including four-phonon scattering, a larger fraction of phonon modes fall into the intermediate regime between the two limits, explaining the enhanced $\kappa_c$ observed in Figure 5a.

\textbf{Figure 5c} displays the four-phonon scattering phase space ($P_4$) decomposed into its three constituent processes: absorption, splitting, and redistribution. The redistribution process dominates across all frequencies, particularly in the 0.5–1.5 THz range where the phonon spectrum exhibits abundant flat bands. These flat band features, originating from the localized vibrations within the isolated TaI$_6$ octahedra and GaI$_4$ tetrahedra, provide ideal conditions for redistribution processes by relaxing momentum conservation and enhancing scattering phase space.

\textbf{Figure 5d} presents the cumulative and differential contributions to $\kappa_p$ and $\kappa_c$ as functions of frequency at 300 K. The differential $\kappa_p$ shows a pronounced peak in the 0–0.5 THz range, indicating that low-frequency acoustic modes with high group velocities but short lifetimes dominate the particle-like transport. In contrast, the differential $\kappa_c$ closely mirrors the profile of the phonon density of states (phDOS), with its maximum occurring in frequency regions where dense phonon bands provide numerous pairs of modes with small frequency differences ($\omega - \omega'$), facilitating wave-like tunneling. Importantly, compared to previously reported ultralow thermal conductivity materials where suppression of $\kappa_p$ is often accompanied by substantial $\kappa_c$ contributions, TaGaI$_8$ maintains a relatively small $\kappa_c$. Microscopically, $\kappa_c$ is promoted by dense manifolds of nearly degenerate modes that enable inter-mode tunneling; however, TaGaI$_8$ exhibits comparatively sparse optical manifolds and larger typical mode-to-mode detunings, which reduce the number of quasi-resonant mode pairs and thus suppress $\kappa_c$\cite{Wu2025a}. Thus, our design strategy based on rotational soft modes and octahedral distortions not only effectively reduces $\kappa_p$ but also avoids inadvertently enhancing $\kappa_c$, yielding a clean suppression of total lattice thermal conductivity.

\section*{Conclusion}
In summary, we have uncovered two physically distinct yet additive mechanisms governing ultralow lattice thermal conductivity in halide perovskites and related octahedral framework materials: halogen-halogen-driven rotational soft modes and static octahedral distortions. Using first-principles calculations on CsPbBr$_3$, we demonstrate that Br-Br interactions induce rotational soft modes that decongest the phonon spectrum, enhance three- and four-phonon scattering phase space, and strongly suppress particle-like thermal conductivity ($\kappa_p$). Independently, static octahedral distortions further reduce $\kappa_p$ by amplifying anharmonicity while leaving wave-like conductivity ($\kappa_c$) intact. Based on these mechanistic insights, we introduce a geometric distortion factor $\rho$ and perform high-throughput screening to identify TaGaI$_8$ with an ultralow $\kappa_L = 0.11$ W/mK at room temperature. This work establishes halogen–halogen interactions as a key design element for rotational soft modes and octahedral distortion as an independent strategy for thermal transport modulation, providing a clear framework---combining octahedral building blocks for soft-mode generation with distortion metrics for anharmonicity enhancement---for discovering novel low-thermal-conductivity materials with octahedral frameworks.

\section*{Numerical methods}
The calculations are implemented using the Vienna Ab Initio Simulation Package (VASP) based on density functional theory (DFT)\cite{Kresse1996} with the projector augmented wave (PAW) method and the revised PBE-GGA exchange-correlation functional for solids (PBEsol)\cite{Perdew2008}. The cutoff energy of the plane wave is set to 500 eV. The energy convergence value between two consecutive steps is set as $10^{-5}$ eV when optimizing atomic positions, and the maximum Hellmann-Feynman force acting on each atom is $10^{-3}$ eV/\r{A}.

For CsPbBr$_3$, to ensure consistent symmetry across all structures investigated in Fig.~2, the pristine reference structure is prepared with a minimal Pb displacement of $10^{-5}$ relative to the lattice constant, breaking the high-symmetry configuration while serving as the baseline for subsequent distortion studies. A $3\times3\times3$ supercell containing 135 atoms is constructed for phonon calculations.

For the newly identified materials (AlSbI$_6$, NbAlI$_8$, NbGaI$_8$, TaAlI$_8$, TaGaI$_8$, and HfI$_4$), supercell sizes are chosen as $2\times2\times2$ and $2\times2\times3$ for AlSbI$_6$.

Ab initio molecular dynamics (AIMD) simulations are performed for each system using the constructed supercells. Each simulation is run for 20 ps with a timestep of 1 fs. The second-order, third-order, and fourth-order interatomic force constants (IFCs) are extracted from the AIMD trajectories using the temperature-dependent effective potential (TDEP) method\cite{Hellman2013}, which naturally incorporates temperature renormalization of phonon frequencies. The interaction cutoff radii are set to 10.0, 6.0, and 5.0 Å for second-, third-, and fourth-order IFCs, respectively.

Using the extracted IFCs, the lattice thermal conductivity is calculated with the ShengBTE package\cite{shengbte2014}. For CsPbBr$_3$, the $\mathbf{q}$-mesh in the first irreducible Brillouin zone is set to $12\times12\times12$. For the newly identified materials, the $\mathbf{q}$-mesh is set to $9\times9\times9$ to ensure convergence. The wave-like thermal conductivity ($\kappa_c$) is calculated using our in-house extension integrated into the ShengBTE package.

\section*{Conflicts of interest}
There are no conflicts to declare.

\section*{Acknowledgements}

This work is supported by Natural Science Foundation of China (Grants No. 12304038, 52206092, 12204402), Outstanding Youth Foundation Project of Jiangsu Province (BK20250035), National Key R\&D Program of China (Grants No. 2024YFF0508900), and Big Data Computing Center of Southeast University and the Center for Fundamental and Interdisciplinary Sciences of Southeast University.



\begin{mcitethebibliography}{41}
\providecommand*\natexlab[1]{#1}
\providecommand*\mciteSetBstSublistMode[1]{}
\providecommand*\mciteSetBstMaxWidthForm[2]{}
\providecommand*\mciteBstWouldAddEndPuncttrue
  {\def\EndOfBibitem{\unskip.}}
\providecommand*\mciteBstWouldAddEndPunctfalse
  {\let\EndOfBibitem\relax}
\providecommand*\mciteSetBstMidEndSepPunct[3]{}
\providecommand*\mciteSetBstSublistLabelBeginEnd[3]{}
\providecommand*\EndOfBibitem{}
\mciteSetBstSublistMode{f}
\mciteSetBstMaxWidthForm{subitem}{(\alph{mcitesubitemcount})}
\mciteSetBstSublistLabelBeginEnd
  {\mcitemaxwidthsubitemform\space}
  {\relax}
  {\relax}

\bibitem[He and Liu(2023)He, and Liu]{He2023}
He,~C.; Liu,~X. The rise of halide perovskite semiconductors. \emph{Light Sci.
  Appl.} \textbf{2023}, \emph{12}\relax
\mciteBstWouldAddEndPuncttrue
\mciteSetBstMidEndSepPunct{\mcitedefaultmidpunct}
{\mcitedefaultendpunct}{\mcitedefaultseppunct}\relax
\EndOfBibitem
\bibitem[Brenner \latin{et~al.}(2016)Brenner, Egger, Kronik, Hodes, and
  Cahen]{Brenner2016}
Brenner,~T.~M.; Egger,~D.~A.; Kronik,~L.; Hodes,~G.; Cahen,~D. Hybrid
  organic—inorganic perovskites: low-cost semiconductors with intriguing
  charge-transport properties. \emph{Nat. Rev. Mater.} \textbf{2016},
  \emph{1}\relax
\mciteBstWouldAddEndPuncttrue
\mciteSetBstMidEndSepPunct{\mcitedefaultmidpunct}
{\mcitedefaultendpunct}{\mcitedefaultseppunct}\relax
\EndOfBibitem
\bibitem[Green \latin{et~al.}(2014)Green, Ho-Baillie, and Snaith]{Green2014}
Green,~M.~A.; Ho-Baillie,~A.; Snaith,~H.~J. The emergence of perovskite solar
  cells. \emph{Nat. Photonics} \textbf{2014}, \emph{8}, 506--514\relax
\mciteBstWouldAddEndPuncttrue
\mciteSetBstMidEndSepPunct{\mcitedefaultmidpunct}
{\mcitedefaultendpunct}{\mcitedefaultseppunct}\relax
\EndOfBibitem
\bibitem[Liu \latin{et~al.}(2020)Liu, Xu, Bai, Jin, Wang, Friend, and
  Gao]{Liu2020b}
Liu,~X.-K.; Xu,~W.; Bai,~S.; Jin,~Y.; Wang,~J.; Friend,~R.~H.; Gao,~F. Metal
  halide perovskites for light-emitting diodes. \emph{Nat. Mater.}
  \textbf{2020}, \emph{20}, 10--21\relax
\mciteBstWouldAddEndPuncttrue
\mciteSetBstMidEndSepPunct{\mcitedefaultmidpunct}
{\mcitedefaultendpunct}{\mcitedefaultseppunct}\relax
\EndOfBibitem
\bibitem[Isikgor \latin{et~al.}(2022)Isikgor, Zhumagali, T.~Merino,
  De~Bastiani, McCulloch, and De~Wolf]{Isikgor2022}
Isikgor,~F.~H.; Zhumagali,~S.; T.~Merino,~L.~V.; De~Bastiani,~M.;
  McCulloch,~I.; De~Wolf,~S. Molecular engineering of contact interfaces for
  high-performance perovskite solar cells. \emph{Nat. Rev. Mater.}
  \textbf{2022}, \emph{8}, 89--108\relax
\mciteBstWouldAddEndPuncttrue
\mciteSetBstMidEndSepPunct{\mcitedefaultmidpunct}
{\mcitedefaultendpunct}{\mcitedefaultseppunct}\relax
\EndOfBibitem
\bibitem[Haque \latin{et~al.}(2019)Haque, Troughton, and Baran]{Haque2019}
Haque,~M.~A.; Troughton,~J.; Baran,~D. Processing‐performance evolution of
  perovskite solar cells: from large grain polycrystalline films to single
  crystals. \emph{Adv. Energy Mater.} \textbf{2019}, \emph{10}\relax
\mciteBstWouldAddEndPuncttrue
\mciteSetBstMidEndSepPunct{\mcitedefaultmidpunct}
{\mcitedefaultendpunct}{\mcitedefaultseppunct}\relax
\EndOfBibitem
\bibitem[Haque \latin{et~al.}(2020)Haque, Kee, Villalva, Ong, and
  Baran]{Haque2020}
Haque,~M.~A.; Kee,~S.; Villalva,~D.~R.; Ong,~W.; Baran,~D. Halide perovskites:
  Thermal transport and prospects for thermoelectricity. \emph{Adv. Sci.}
  \textbf{2020}, \emph{7}, 1903389\relax
\mciteBstWouldAddEndPuncttrue
\mciteSetBstMidEndSepPunct{\mcitedefaultmidpunct}
{\mcitedefaultendpunct}{\mcitedefaultseppunct}\relax
\EndOfBibitem
\bibitem[Liang \latin{et~al.}(2017)Liang, Liu, and Jin]{Liang2017}
Liang,~J.; Liu,~J.; Jin,~Z. All‐inorganic halide perovskites for
  optoelectronics: Progress and prospects. \emph{Solar RRL} \textbf{2017},
  \emph{1}, 1700086\relax
\mciteBstWouldAddEndPuncttrue
\mciteSetBstMidEndSepPunct{\mcitedefaultmidpunct}
{\mcitedefaultendpunct}{\mcitedefaultseppunct}\relax
\EndOfBibitem
\bibitem[Yuan \latin{et~al.}(2017)Yuan, Li, Wang, Xing, Gruverman, and
  Huang]{Yuan2017}
Yuan,~Y.; Li,~T.; Wang,~Q.; Xing,~J.; Gruverman,~A.; Huang,~J. Anomalous
  photovoltaic effect in organic-inorganic hybrid perovskite solar cells.
  \emph{Sci. Adv.} \textbf{2017}, \emph{3}, e1602164\relax
\mciteBstWouldAddEndPuncttrue
\mciteSetBstMidEndSepPunct{\mcitedefaultmidpunct}
{\mcitedefaultendpunct}{\mcitedefaultseppunct}\relax
\EndOfBibitem
\bibitem[Ha \latin{et~al.}(2015)Ha, Shen, Zhang, and Xiong]{Ha2015}
Ha,~S.-T.; Shen,~C.; Zhang,~J.; Xiong,~Q. Laser cooling of organic–inorganic
  lead halide perovskites. \emph{Nat. Photonics} \textbf{2015}, \emph{10},
  115--121\relax
\mciteBstWouldAddEndPuncttrue
\mciteSetBstMidEndSepPunct{\mcitedefaultmidpunct}
{\mcitedefaultendpunct}{\mcitedefaultseppunct}\relax
\EndOfBibitem
\bibitem[Pazos-Outón \latin{et~al.}(2016)Pazos-Outón, Szumilo, Lamboll,
  Richter, Crespo-Quesada, Abdi-Jalebi, Beeson, Vrućinić, Alsari, Snaith,
  Ehrler, Friend, and Deschler]{Pazos-Outon2016}
Pazos-Outón,~L.~M.; Szumilo,~M.; Lamboll,~R.; Richter,~J.~M.;
  Crespo-Quesada,~M.; Abdi-Jalebi,~M.; Beeson,~H.~J.; Vrućinić,~M.;
  Alsari,~M.; Snaith,~H.~J.; Ehrler,~B.; Friend,~R.~H.; Deschler,~F. Photon
  recycling in lead iodide perovskite solar cells. \emph{Science}
  \textbf{2016}, \emph{351}, 1430--1433\relax
\mciteBstWouldAddEndPuncttrue
\mciteSetBstMidEndSepPunct{\mcitedefaultmidpunct}
{\mcitedefaultendpunct}{\mcitedefaultseppunct}\relax
\EndOfBibitem
\bibitem[Lee \latin{et~al.}(2017)Lee, Li, Wong, Zhang, Lai, Yu, Kong, Lin,
  Urban, Grossman, and Yang]{Lee2017}
Lee,~W.; Li,~H.; Wong,~A.~B.; Zhang,~D.; Lai,~M.; Yu,~Y.; Kong,~Q.; Lin,~E.;
  Urban,~J.~J.; Grossman,~J.~C.; Yang,~P. Ultralow thermal conductivity in
  all-inorganic halide perovskites. \emph{Proc. Natl. Acad. Sci. U.S.A.}
  \textbf{2017}, \emph{114}, 8693--8697\relax
\mciteBstWouldAddEndPuncttrue
\mciteSetBstMidEndSepPunct{\mcitedefaultmidpunct}
{\mcitedefaultendpunct}{\mcitedefaultseppunct}\relax
\EndOfBibitem
\bibitem[Zheng \latin{et~al.}(2024)Zheng, Lin, Lin, Hautier, Guo, and
  Huang]{Zheng2024}
Zheng,~J.; Lin,~C.; Lin,~C.; Hautier,~G.; Guo,~R.; Huang,~B. Unravelling
  ultralow thermal conductivity in perovskite Cs2AgBiBr6: dominant wave-like
  phonon tunnelling and strong anharmonicity. \emph{NPJ Comput. Mater.}
  \textbf{2024}, \emph{10}\relax
\mciteBstWouldAddEndPuncttrue
\mciteSetBstMidEndSepPunct{\mcitedefaultmidpunct}
{\mcitedefaultendpunct}{\mcitedefaultseppunct}\relax
\EndOfBibitem
\bibitem[Acharyya \latin{et~al.}(2022)Acharyya, Ghosh, Pal, Rana, Dutta, Swain,
  Etter, Soni, Waghmare, and Biswas]{Acharyya2022}
Acharyya,~P.; Ghosh,~T.; Pal,~K.; Rana,~K.~S.; Dutta,~M.; Swain,~D.; Etter,~M.;
  Soni,~A.; Waghmare,~U.~V.; Biswas,~K. Glassy thermal conductivity in
  Cs3Bi2I6Cl3 single crystal. \emph{Nat. Commun.} \textbf{2022}, \emph{13},
  5053\relax
\mciteBstWouldAddEndPuncttrue
\mciteSetBstMidEndSepPunct{\mcitedefaultmidpunct}
{\mcitedefaultendpunct}{\mcitedefaultseppunct}\relax
\EndOfBibitem
\bibitem[Wu \latin{et~al.}(2024)Wu, Huang, Ji, Ji, Ding, and Zhou]{Wu2024}
Wu,~Y.; Huang,~A.; Ji,~L.; Ji,~J.; Ding,~Y.; Zhou,~L. Origin of Intrinsically
  Low Lattice Thermal Conductivity in Solids. \emph{J. Phys. Chem. Lett.}
  \textbf{2024}, \emph{15}, 11525--11537\relax
\mciteBstWouldAddEndPuncttrue
\mciteSetBstMidEndSepPunct{\mcitedefaultmidpunct}
{\mcitedefaultendpunct}{\mcitedefaultseppunct}\relax
\EndOfBibitem
\bibitem[Acharyya \latin{et~al.}(2023)Acharyya, Pal, Ahad, Sarkar, Rana, Dutta,
  Soni, Waghmare, and Biswas]{Acharyya2023}
Acharyya,~P.; Pal,~K.; Ahad,~A.; Sarkar,~D.; Rana,~K.~S.; Dutta,~M.; Soni,~A.;
  Waghmare,~U.~V.; Biswas,~K. Extended antibonding states and phonon
  localization induce ultralow thermal conductivity in low dimensional metal
  halide. \emph{Adv Funct. Mater.} \textbf{2023}, \emph{33}, 2304607\relax
\mciteBstWouldAddEndPuncttrue
\mciteSetBstMidEndSepPunct{\mcitedefaultmidpunct}
{\mcitedefaultendpunct}{\mcitedefaultseppunct}\relax
\EndOfBibitem
\bibitem[Hu \latin{et~al.}(2021)Hu, Ren, Djurišić, and Rogach]{Hu2021}
Hu,~S.; Ren,~Z.; Djurišić,~A.~B.; Rogach,~A.~L. Metal halide perovskites as
  emerging thermoelectric materials. \emph{ACS Energy Lett.} \textbf{2021},
  \emph{6}, 3882--3905\relax
\mciteBstWouldAddEndPuncttrue
\mciteSetBstMidEndSepPunct{\mcitedefaultmidpunct}
{\mcitedefaultendpunct}{\mcitedefaultseppunct}\relax
\EndOfBibitem
\bibitem[Liu \latin{et~al.}(2019)Liu, Zhao, Li, Liu, Liscio, Milita, Schroeder,
  and Fenwick]{Liu2019a}
Liu,~T.; Zhao,~X.; Li,~J.; Liu,~Z.; Liscio,~F.; Milita,~S.; Schroeder,~B.~C.;
  Fenwick,~O. Enhanced control of self-doping in halide perovskites for
  improved thermoelectric performance. \emph{Nat. Commun.} \textbf{2019},
  \emph{10}, 5750\relax
\mciteBstWouldAddEndPuncttrue
\mciteSetBstMidEndSepPunct{\mcitedefaultmidpunct}
{\mcitedefaultendpunct}{\mcitedefaultseppunct}\relax
\EndOfBibitem
\bibitem[Xie \latin{et~al.}(2020)Xie, Hao, Bao, Slade, Snyder, Wolverton, and
  Kanatzidis]{Xie2020}
Xie,~H.; Hao,~S.; Bao,~J.; Slade,~T.~J.; Snyder,~G.~J.; Wolverton,~C.;
  Kanatzidis,~M.~G. All-inorganic halide perovskites as potential
  thermoelectric materials: Dynamic cation off-centering induces ultralow
  thermal conductivity. \emph{J. Am. Chem. Soc} \textbf{2020}, \emph{142},
  9553--9563\relax
\mciteBstWouldAddEndPuncttrue
\mciteSetBstMidEndSepPunct{\mcitedefaultmidpunct}
{\mcitedefaultendpunct}{\mcitedefaultseppunct}\relax
\EndOfBibitem
\bibitem[Haque \latin{et~al.}(2020)Haque, Hernandez, Davaasuren, Villalva,
  Troughton, and Baran]{Haque2020b}
Haque,~M.~A.; Hernandez,~L.~H.; Davaasuren,~B.; Villalva,~D.~R.; Troughton,~J.;
  Baran,~D. Tuning the Thermoelectric Performance of Hybrid Tin Perovskites by
  Air Treatment. \emph{Adv. Energ. Sust. Res.} \textbf{2020}, \emph{1}\relax
\mciteBstWouldAddEndPuncttrue
\mciteSetBstMidEndSepPunct{\mcitedefaultmidpunct}
{\mcitedefaultendpunct}{\mcitedefaultseppunct}\relax
\EndOfBibitem
\bibitem[Wang \latin{et~al.}(2022)Wang, Xu, Li, Li, Liu, Li, Shan, Li, Shi, and
  Kyaw]{Wang2022}
Wang,~T.; Xu,~X.; Li,~W.; Li,~Y.; Liu,~Q.; Li,~C.; Shan,~C.; Li,~G.; Shi,~T.;
  Kyaw,~A. K.~K. Simultaneous Enhancement of Thermoelectric Power Factor and
  Phase Stability of Tin-Based Perovskites by Organic Cation Doping. \emph{ACS
  Appl. Energy Mater.} \textbf{2022}, \emph{5}, 11191--11199\relax
\mciteBstWouldAddEndPuncttrue
\mciteSetBstMidEndSepPunct{\mcitedefaultmidpunct}
{\mcitedefaultendpunct}{\mcitedefaultseppunct}\relax
\EndOfBibitem
\bibitem[Bechtel and Van~der Ven(2018)Bechtel, and Van~der Ven]{Bechtel2018}
Bechtel,~J.~S.; Van~der Ven,~A. Octahedral tilting instabilities in inorganic
  halide perovskites. \emph{Phys. Rev. Mater.} \textbf{2018}, \emph{2}\relax
\mciteBstWouldAddEndPuncttrue
\mciteSetBstMidEndSepPunct{\mcitedefaultmidpunct}
{\mcitedefaultendpunct}{\mcitedefaultseppunct}\relax
\EndOfBibitem
\bibitem[Wang \latin{et~al.}(2025)Wang, Wang, Doherty, Stranks, Gao, and
  Yang]{Wang2025}
Wang,~Y.; Wang,~Y.; Doherty,~T. A.~S.; Stranks,~S.~D.; Gao,~F.; Yang,~D.
  Octahedral units in halide perovskites. \emph{Nat. Rev. Chem.} \textbf{2025},
  \emph{9}, 261--277\relax
\mciteBstWouldAddEndPuncttrue
\mciteSetBstMidEndSepPunct{\mcitedefaultmidpunct}
{\mcitedefaultendpunct}{\mcitedefaultseppunct}\relax
\EndOfBibitem
\bibitem[Acharyya \latin{et~al.}(2020)Acharyya, Ghosh, Pal, Kundu, Singh~Rana,
  Pandey, Soni, Waghmare, and Biswas]{Acharyya2020}
Acharyya,~P.; Ghosh,~T.; Pal,~K.; Kundu,~K.; Singh~Rana,~K.; Pandey,~J.;
  Soni,~A.; Waghmare,~U.~V.; Biswas,~K. Intrinsically Ultralow Thermal
  Conductivity in Ruddlesden–Popper 2D Perovskite Cs2PbI2Cl2: Localized
  Anharmonic Vibrations and Dynamic Octahedral Distortions. \emph{J. Am. Chem.
  Soc} \textbf{2020}, \emph{142}, 15595--15603\relax
\mciteBstWouldAddEndPuncttrue
\mciteSetBstMidEndSepPunct{\mcitedefaultmidpunct}
{\mcitedefaultendpunct}{\mcitedefaultseppunct}\relax
\EndOfBibitem
\bibitem[Xie and Zhao(2024)Xie, and Zhao]{Xie2024}
Xie,~H.; Zhao,~L.-D. Origin of off-centering effect and the influence on heat
  transport in thermoelectrics. \emph{Materials Futures} \textbf{2024},
  \emph{3}, 013501\relax
\mciteBstWouldAddEndPuncttrue
\mciteSetBstMidEndSepPunct{\mcitedefaultmidpunct}
{\mcitedefaultendpunct}{\mcitedefaultseppunct}\relax
\EndOfBibitem
\bibitem[Christensen \latin{et~al.}(2008)Christensen, Abrahamsen, Christensen,
  Juranyi, Andersen, Lefmann, Andreasson, Bahl, and Iversen]{Christensen2008}
Christensen,~M.; Abrahamsen,~A.~B.; Christensen,~N.~B.; Juranyi,~F.;
  Andersen,~N.~H.; Lefmann,~K.; Andreasson,~J.; Bahl,~C. R.~H.; Iversen,~B.~B.
  Avoided crossing of rattler modes in thermoelectric materials. \emph{Nat.
  Mater.} \textbf{2008}, \emph{7}, 811--815\relax
\mciteBstWouldAddEndPuncttrue
\mciteSetBstMidEndSepPunct{\mcitedefaultmidpunct}
{\mcitedefaultendpunct}{\mcitedefaultseppunct}\relax
\EndOfBibitem
\bibitem[Tadano \latin{et~al.}(2015)Tadano, Gohda, and Tsuneyuki]{Tadano2015}
Tadano,~T.; Gohda,~Y.; Tsuneyuki,~S. Impact of rattlers on thermal conductivity
  of a thermoelectric clathrate: A first-principles study. \emph{Phys. Rev.
  Lett.} \textbf{2015}, \emph{114}\relax
\mciteBstWouldAddEndPuncttrue
\mciteSetBstMidEndSepPunct{\mcitedefaultmidpunct}
{\mcitedefaultendpunct}{\mcitedefaultseppunct}\relax
\EndOfBibitem
\bibitem[Thakur and Giri(2023)Thakur, and Giri]{Thakur2023}
Thakur,~S.; Giri,~A. Origin of ultralow thermal conductivity in metal halide
  perovskites. \emph{ACS Appl. Mater. Interfaces} \textbf{2023}, \emph{15},
  26755--26765\relax
\mciteBstWouldAddEndPuncttrue
\mciteSetBstMidEndSepPunct{\mcitedefaultmidpunct}
{\mcitedefaultendpunct}{\mcitedefaultseppunct}\relax
\EndOfBibitem
\bibitem[Wu \latin{et~al.}(2025)Wu, Ji, Zeng, Chen, Liu, and Zhou]{Wu2025}
Wu,~Y.; Ji,~L.; Zeng,~S.; Chen,~Y.; Liu,~C.; Zhou,~L. Weak host interactions
  induced thermal transport properties of metal halide perovskites deviating
  from the rattling model. \emph{J. Phys. Chem. Lett.} \textbf{2025},
  \emph{16}, 10035--10041\relax
\mciteBstWouldAddEndPuncttrue
\mciteSetBstMidEndSepPunct{\mcitedefaultmidpunct}
{\mcitedefaultendpunct}{\mcitedefaultseppunct}\relax
\EndOfBibitem
\bibitem[Pandey \latin{et~al.}(2022)Pandey, Du, Parker, and
  Lindsay]{PANDEY2022}
Pandey,~T.; Du,~M.-H.; Parker,~D.~S.; Lindsay,~L. Origin of ultralow phonon
  transport and strong anharmonicity in lead-free halide perovskites.
  \emph{Mater. Today Phys.} \textbf{2022}, \emph{28}, 100881\relax
\mciteBstWouldAddEndPuncttrue
\mciteSetBstMidEndSepPunct{\mcitedefaultmidpunct}
{\mcitedefaultendpunct}{\mcitedefaultseppunct}\relax
\EndOfBibitem
\bibitem[Zeng \latin{et~al.}(2025)Zeng, Fan, Simoncelli, Chen, Liang, Chen,
  Thornton, and Cheng]{Zeng2025}
Zeng,~Z.; Fan,~Z.; Simoncelli,~M.; Chen,~C.; Liang,~T.; Chen,~Y.; Thornton,~G.;
  Cheng,~B. Lattice distortion leads to glassy thermal transport in crystalline
  Cs3Bi2I6Cl3. \emph{Proc. Natl. Acad. Sci. U.S.A.} \textbf{2025},
  \emph{122}\relax
\mciteBstWouldAddEndPuncttrue
\mciteSetBstMidEndSepPunct{\mcitedefaultmidpunct}
{\mcitedefaultendpunct}{\mcitedefaultseppunct}\relax
\EndOfBibitem
\bibitem[Di~Lucente \latin{et~al.}(2023)Di~Lucente, Simoncelli, and
  Marzari]{DiLucente2023}
Di~Lucente,~E.; Simoncelli,~M.; Marzari,~N. Crossover from Boltzmann to Wigner
  thermal transport in thermoelectric skutterudites. \emph{Phys. Rev. Res.}
  \textbf{2023}, \emph{5}\relax
\mciteBstWouldAddEndPuncttrue
\mciteSetBstMidEndSepPunct{\mcitedefaultmidpunct}
{\mcitedefaultendpunct}{\mcitedefaultseppunct}\relax
\EndOfBibitem
\bibitem[Simoncelli \latin{et~al.}(2019)Simoncelli, Marzari, and
  Mauri]{Simoncelli2019}
Simoncelli,~M.; Marzari,~N.; Mauri,~F. Unified theory of thermal transport in
  crystals and glasses. \emph{Nat. Phys.} \textbf{2019}, \emph{15},
  809--813\relax
\mciteBstWouldAddEndPuncttrue
\mciteSetBstMidEndSepPunct{\mcitedefaultmidpunct}
{\mcitedefaultendpunct}{\mcitedefaultseppunct}\relax
\EndOfBibitem
\bibitem[Wu \latin{et~al.}(2023)Wu, Chen, Fang, Ding, Li, Xue, Shao, Zhang, and
  Zhou]{Wu2023b}
Wu,~Y.; Chen,~Y.; Fang,~Z.; Ding,~Y.; Li,~Q.; Xue,~K.; Shao,~H.; Zhang,~H.;
  Zhou,~L. Ultralow lattice thermal transport and considerable wave-like phonon
  tunneling in chalcogenide perovskite BaZrS3. \emph{J. Phys. Chem. Lett.}
  \textbf{2023}, \emph{14}, 11465--11473\relax
\mciteBstWouldAddEndPuncttrue
\mciteSetBstMidEndSepPunct{\mcitedefaultmidpunct}
{\mcitedefaultendpunct}{\mcitedefaultseppunct}\relax
\EndOfBibitem
\bibitem[Wu \latin{et~al.}(2025)Wu, Chen, Zeng, Zhou, and Liu]{Wu2025b}
Wu,~Y.; Chen,~Y.; Zeng,~S.; Zhou,~L.; Liu,~C. High n-type thermoelectric
  performance due to anisotropic charge-phonon transport in CuBiSCl2.
  \emph{Phys. Rev. B} \textbf{2025}, \emph{112}, 054306\relax
\mciteBstWouldAddEndPuncttrue
\mciteSetBstMidEndSepPunct{\mcitedefaultmidpunct}
{\mcitedefaultendpunct}{\mcitedefaultseppunct}\relax
\EndOfBibitem
\bibitem[Wu \latin{et~al.}(2025)Wu, Chen, Zeng, Li, Zhang, Zhou, Wei, and
  Liu]{Wu2025a}
Wu,~Y.; Chen,~Y.; Zeng,~S.; Li,~G.; Zhang,~H.; Zhou,~L.; Wei,~S.-H.; Liu,~C.
  Pushing the Thermal Conductivity Limit by Decoupling Dual-Channel Phonon
  Transport in Crystals. \emph{Nano Lett.} \textbf{2025}, \emph{26},
  1632--1638\relax
\mciteBstWouldAddEndPuncttrue
\mciteSetBstMidEndSepPunct{\mcitedefaultmidpunct}
{\mcitedefaultendpunct}{\mcitedefaultseppunct}\relax
\EndOfBibitem
\bibitem[Kresse and Furthmuller({1996})Kresse, and Furthmuller]{Kresse1996}
Kresse,~G.; Furthmuller,~J. {Efficient iterative schemes for ab initio
  total-energy calculations using a plane-wave basis set}. \emph{{Phys. Rev.
  B}} \textbf{{1996}}, \emph{{54}}, {11169--11186}\relax
\mciteBstWouldAddEndPuncttrue
\mciteSetBstMidEndSepPunct{\mcitedefaultmidpunct}
{\mcitedefaultendpunct}{\mcitedefaultseppunct}\relax
\EndOfBibitem
\bibitem[Perdew \latin{et~al.}(2008)Perdew, Ruzsinszky, Csonka, Vydrov,
  Scuseria, Constantin, Zhou, and Burke]{Perdew2008}
Perdew,~J.~P.; Ruzsinszky,~A.; Csonka,~G.~I.; Vydrov,~O.~A.; Scuseria,~G.~E.;
  Constantin,~L.~A.; Zhou,~X.; Burke,~K. Restoring the density-gradient
  expansion for exchange in solids and surfaces. \emph{Phys. Rev. Lett.}
  \textbf{2008}, \emph{100}, 136406\relax
\mciteBstWouldAddEndPuncttrue
\mciteSetBstMidEndSepPunct{\mcitedefaultmidpunct}
{\mcitedefaultendpunct}{\mcitedefaultseppunct}\relax
\EndOfBibitem
\bibitem[Hellman and Abrikosov(2013)Hellman, and Abrikosov]{Hellman2013}
Hellman,~O.; Abrikosov,~I.~A. Temperature-dependent effective third-order
  interatomic force constants from first principles. \emph{Phys. Rev. B}
  \textbf{2013}, \emph{88}, 144301\relax
\mciteBstWouldAddEndPuncttrue
\mciteSetBstMidEndSepPunct{\mcitedefaultmidpunct}
{\mcitedefaultendpunct}{\mcitedefaultseppunct}\relax
\EndOfBibitem
\bibitem[Li \latin{et~al.}({2014})Li, Carrete, Katcho, and Mingo]{shengbte2014}
Li,~W.; Carrete,~J.; Katcho,~N.~A.; Mingo,~N. {ShengBTE: A solver of the
  Boltzmann transport equation for phonons}. \emph{Comput. Phys. Commun.}
  \textbf{{2014}}, \emph{{185}}, {1747--1758}\relax
\mciteBstWouldAddEndPuncttrue
\mciteSetBstMidEndSepPunct{\mcitedefaultmidpunct}
{\mcitedefaultendpunct}{\mcitedefaultseppunct}\relax
\EndOfBibitem
\end{mcitethebibliography}

\providecommand{\latin}[1]{#1}
\makeatletter
\providecommand{\doi}
  {\begingroup\let\do\@makeother\dospecials
  \catcode`\{=1 \catcode`\}=2 \doi@aux}
\providecommand{\doi@aux}[1]{\endgroup\texttt{#1}}
\makeatother
\providecommand*\mcitethebibliography{\thebibliography}
\csname @ifundefined\endcsname{endmcitethebibliography}
  {\let\endmcitethebibliography\endthebibliography}{}

\end{document}